\pdfoutput=1
\documentclass[letter,traditabstract,longauth]{myaa}
\usepackage{natbib}
\usepackage{lscape}
\usepackage[section]{placeins}
\usepackage{wasysym}
\usepackage{graphics}
\usepackage{graphicx}
\usepackage{txfonts}
\usepackage{color}
\usepackage{times}
\usepackage{mathtools}
\usepackage[T1]{fontenc}
\newcommand{\changefont}[3]{\fontfamily{#1}\fontseries{#2}\fontshape{#3}\selectfont}
\newfont{\vcap}{cmssdc10 scaled 1000}
\newfont{\lcap}{cmssdc10 scaled 1100}
\newfont{\nlx}{cmssdc10 scaled 900}
\newcommand{\rem}[1]{\nlx {#1}\normalfont}
\newcommand{\sbb}{mag/$\sq\arcsec$}

\def\rr{{\sl R}$^{\star}$}
\def\lyc{{\sl Ly}$_{\rm c}$}
\def\pAGBp{pAGB photoinization}
\def\ige{\changefont{cmtt}{m}{it}ne\rm}
\def\wism{\changefont{cmtt}{m}{it}wim\rm}
\def\plf{\changefont{cmtt}{m}{it}plf\rm}
\def\ff{\changefont{cmtt}{m}{it}f\rm}
\def\ha{H$\alpha$}

\def\sstar{$\Sigma_{\star}$}
\def\lo3hb{$\log$([O\,{\sc iii}]/H$\beta$)}
\def\ln2ha{$\log$([N\,{\sc ii}]/H$\alpha$)}
\def\tauha{$\tau$}
\def\tauha_ext{$\tau$}
\def\rPetr{$r_{\mathrm{p}}$}
\def\pAGBmin{{\sc ew}$_{\star}^-$}
\def\pAGBmax{{\sc ew}$_{\star}^+$}
\def\sisp{\nlx sisp\rm} 
\def\iaa{\nlx isan\rm} 
\newcommand {\aga} {\ {\raise-.5ex\hbox{$\buildrel>\over\sim$}}\ }
\newcommand {\ala} {\ {\raise-.5ex\hbox{$\buildrel<\over\sim$}}\ } 
\begin{document}
%
% :::::::::::::::::::::::::::::::::::::::::::::::::::::::::::::::::::::::::::::::::
\titlerunning{Nebular emission and the Lyman continuum photon escape fraction in CALIFA early-type galaxies}
\authorrunning{P. Papaderos}
\title{Nebular emission and the Lyman continuum photon escape fraction in CALIFA early-type galaxies
\thanks{Based on observations collected at the Centro Astron\'omico
Hispano Alem\'an (CAHA) at Calar Alto, operated jointly by the Max-Planck-Institut 
f\"ur Astronomie (MPIA) and the Instituto de Astrof\'isica de
Andaluc\'ia (CSIC)}}
% \subtitle{...}

\author{
P. Papaderos\inst{\ref{inst1}}
\and J.M. Gomes\inst{\ref{inst1}}
\and J.M. V\'{\i}lchez\inst{\ref{inst2}}
\and C. Kehrig\inst{\ref{inst2}}
\and M.D. Lehnert\inst{\ref{inst3}}
\and B. Ziegler\inst{\ref{inst4}}
\and S.~F.~S\'anchez\inst{\ref{inst5},\ref{inst2}}
\and B.~Husemann\inst{\ref{inst6}}
\and A.~Monreal-Ibero\inst{\ref{inst2}}
\and R.~Garc{\'\i}a-Benito\inst{\ref{inst2}}
\and J.~Bland-Hawthorn\inst{\ref{inst7}}
\and C.~Coritjo\inst{\ref{inst2}}
\and A.~de~Lorenzo-C{\'a}ceres\inst{\ref{inst8}}
\and A.~del~Olmo\inst{\ref{inst2}}
\and J.~Falc\'on-Barroso\inst{\ref{IAC},\ref{UnivLaguna}}
\and L.~Galbany\inst{\ref{inst9}}
\and J.~Iglesias-P\'aramo\inst{\ref{inst2},\ref{inst5}}
\and \'A.R. L\'opez-S\'anchez\inst{\ref{inst10},\ref{inst11}}
\and I. Marquez\inst{\ref{inst2}}
\and M.~Moll{\'a}\inst{\ref{inst12}} 
\and D.~Mast\inst{\ref{inst5},\ref{inst2}}
\and G.~van~de~Ven\inst{\ref{inst13}}
\and L.~Wisotzki\inst{\ref{inst6}}
\and the CALIFA collaboration
}
%
%\offprints{papaderos@astro.up.pt}

\institute{
% Institute 1
Centro de Astrof{\'\i}sica and Faculdade de Ci\^encias, Universidade do Porto, Rua das Estrelas, 4150-762 Porto, Portugal\label{inst1}
\and
% Institute 2
Instituto de Astrof\'isica de Andaluc\'ia (CSIC), Glorieta de la Astronom\'{\i}a s/n Aptdo. 3004, E18080-Granada, Spain\label{inst2}
\and
% Institute 3
GEPI, Observatoire de Paris, UMR 8111, CNRS, Université Paris Diderot, 5 place Jules Janssen, 92190 Meudon, France\label{inst3}
\and
% Institute 4
University of Vienna, T\"{u}rkenschanzstrasse 17, 1180 Vienna, Austria\label{inst4}
\and
% Institute 5
Centro Astron\'omico Hispano Alem\'an de Calar Alto (CSIC-MPG), C/ Jes\'us Durb\'an Rem\'on 2-2, E-4004 Almer\'ia, Spain\label{inst5}
\and
% Institute 6
Leibniz-Institut f\"ur Astrophysik Potsdam (AIP), An der Sternwarte 16, D-14482 Potsdam, Germany\label{inst6}
\and
% Institute 7
Sydney Institute for Astronomy, University of Sydney, NSW 2006, Australia\label{inst7}
\and
% Institute 8
School of Physics and Astronomy, University of St Andrews, North Haugh, St
Andrews, KY16 9SS, UK (SUPA)\label{inst8}
\and 
% IAC
Instituto de Astrof\'isica de Canarias, V\'ia L\'actea s/n, La Laguna, Tenerife, Spain\label{IAC}
\and
% Univ. Laguna
Departamento de Astrof\'isica, Universidad de La Laguna, E-38205 La Laguna, Tenerife, Spain\label{UnivLaguna}
% Institute 9
\and
CENTRA - Centro Multidisciplinar de Astrof\'isica, Instituto Superior T\'ecnico, Av. Rovisco Pais 1, 1049-001 Lisbon, Portugal\label{inst9}
\and
% Institute 10
Australian Astronomical Observatory, PO Box 915, North Ryde, NSW 1670, Australia\label{inst10}
\and
% Institute 11
Department of Physics and Astronomy, Macquarie University, NSW 2109, Australia\label{inst11}
\and
% Institute 12
Departamento de Investigaci\'{o}n B\'{a}sica, CIEMAT, Avda. Complutense 40, E-28040 Madrid, Spain\label{inst12}
\and
% Institute 10
Max-Planck-Institut f\"ur Astronomie, K\"onigstuhl 17, D-69117 Heidelberg, Germany\label{inst13}
}
\date{Received 11 April 2013 / Accepted 2 June 2013}
\abstract{
We use deep integral field spectroscopy data from the CALIFA survey to study the warm interstellar medium (\wism) 
over the entire extent and optical spectral range of 32 nearby early-type galaxies (ETGs).
We find that faint nebular emission is extended in all cases, and its surface brightness 
decreases roughly as $\propto r^{-{\alpha}}$. The large standard deviation in the derived 
$\alpha$ (1.09$\pm$0.67) argues against a \emph{universal} power-law index for the radial 
drop-off of nebular emission in ETGs.
Judging from the properties of their extranuclear component, our sample ETGs 
span a broad, continuous sequence with respect to their $\alpha$, \ha\ equivalent width (EW) 
and Lyman continuum (\lyc) photon leakage fraction (\plf).
We propose a tentative subdivision into two groups:
Type~i ETGs are characterized by rather steep \ha\ profiles ($\alpha\simeq 1.4$), 
comparatively large (\aga1~\AA), nearly radially constant EWs, and \plf$\simeq$0. 
Photoionization by post-AGB stars appears to be the main driver of extended nebular emission 
in these systems, with nonthermal sources being potentially important only in their nuclei.
Typical properties of type~ii ETGs are shallower \ha\ profiles ($\alpha\simeq 0.8$), 
very low (\ala0.5~\AA) EWs with positive radial gradients, and a mean 
\plf\aga0.7, rising to \aga0.9 in their centers. 
Such properties point to a low, and inwardly decreasing \wism\ density and/or volume filling factor.
We argue that, because of extensive \lyc\ photon leakage, emission-line luminosities and EWs 
are reduced in type~ii ETG nuclei by at least one order of magnitude.
Consequently, the line weakness of these ETGs is by itself no compelling evidence 
for their containing merely ``weak'' (sub-Eddington accreting) active galactic nuclei (AGN).
In fact, \lyc\ photon escape, which has heretofore not been considered, may constitute
a key element in understanding why many ETGs with prominent signatures of AGN
activity in radio continuum and/or X-ray wavelengths show only faint emission
lines and weak signatures of AGN activity in their optical spectra. 
The \lyc\ photon escape, in conjunction with dilution of nuclear EWs by line-of-sight integration 
through a \emph{triaxial} stellar host, can systematically impede detection of AGN in gas-poor 
galaxy spheroids through optical emission-line spectroscopy, thereby leading to an observational bias.
We further find that type~i\&ii ETGs differ little (\ala0.4~dex) in their mean 
BPT line ratios, which in both cases are characteristic of LINERs and are, within their uncertainties, 
almost radius-independent. 
This potentially hints at a degeneracy of the projected, luminosity-weighted BPT ratios in the LINER regime, 
for the specific 3D properties of the \wism\ and the ionizing photon field in ETGs.
} 
\keywords{galaxies: elliptical and lenticular, cD - galaxies: nuclei - galaxies: ISM}
\maketitle

% ========================================================================
\section{Introduction \label{intro}}
% ========================================================================
Even though the presence of faint nebular emission (\ige) in the nuclei of many early-type galaxies (ETGs) 
has long been established observationally \citep[e.g.,][hereafter K12]{Phillips1986,sar06,sar10,ani10,Kehrig2012},
the nature of the dominant excitation mechanism of the warm interstellar medium (\wism)
in these systems remains uncertain.
The {\sl low-ionization nuclear emission-line region} (LINER) emission-line ratios,
as a typical property of ETG nuclei, have prompted various interpretations
\citep[see, e.g., K12,][]{YanBlanton2012},
including low-accretion rate active galactic nuclei \citep[AGN; e.g.,][]{Ho1999}, 
fast shocks \citep[e.g.][]{dop95}, and hot, evolved ($\geq 10^8$ yr) post-AGB (pAGB) 
stars \citep[e.g.,][]{tri91,bin94,sta08}.
Since each of these mechanisms is tied to distinct and testable expectations on the 
2D properties of the \wism, the limited spatial coverage of previous  
single-aperture and longslit spectroscopic studies has been an important 
obstacle to any conclusive discrimination between them.
Spatially resolved integral field spectroscopy (IFS) over the entire extent of 
ETGs offers an essential advantage in this respect and promises key observational constraints
toward the resolution of this longstanding debate.

This Letter gives a brief summary of our results from an ongoing study of
32 ETGs, which were mapped with deep IFS over their entire extent and optical spectral range 
with the goal of gaining deeper insight into the 2D properties of their \wism. 
A detailed discussion of individual objects and our methodology will be given   
in Gomes et al. (2013, in prep.; hereafter G13) and subsequent publications of this series.
This study is based on low-spectral-resolution ($R\sim 850$) IFS cubes for 
20 E and 12 S0 nearby ($<$150~Mpc) galaxies from the {\sl Calar Alto Legacy Integral Field Area}
(CALIFA) survey \citep[][Walcher et al. 2013, in prep.]{Sanchez2012}. 
These data are being made accessible to the community in a fully reduced and well-documented 
format \citep{Husemann2013} through successive data releases. 

% ==========================================================================
\section{Methodology and results \label{meth}}
% ==========================================================================
The CALIFA data cubes were processed with the \rem{Porto3D} pipeline
(see K12 and G13 for details), which, among various other tasks, permits 
spaxel-by-spaxel spectral fitting of the stellar component with the population 
synthesis code {\sc starlight} \citep[][]{cid05} and subsequent determination of 
emission line fluxes and their uncertainties from the pure
emission-line spectrum (i.e. the observed spectrum after subtraction of the 
best-fitting synthetic stellar model).
For each ETG, typically $\sim$1600 to $\sim$3400 individual spectra with a S/N$\geq$30 at 5150~\AA\ 
were extracted and modeled in the spectral range 4000--6800~\AA\ using both \citet[][hereafter BC]{bru03} and 
MILES \citep{san06,vaz10} simple-stellar population (SSP) libraries, which  
comprise 34 ages between 5 Myr and 13 Gyr for three metallicities 
(0.008, 0.019, and 0.03), i.e., 102 elements each.
After full analysis and cross-inspection of the relevant output from the BC- and MILES-based models,
the emission-line maps for each ETG were error-weighted and averaged spaxel-by-spaxel 
to reduce uncertainties.

An extra module in \rem{Porto3D} permits computation of the Lyman continuum (\lyc) ionizing photon rate 
corresponding to the best-fitting set of BC SSPs for each spaxel. 
The \lyc\ output is then converted into Balmer line luminosities 
assuming case~B recombination for an electron temperature and density 
of $10^4$~K and 100 cm$^{-3}$, respectively.
The same module computes the distance-independent $\tau$ ratio of 
the \ha\ luminosity predicted from pAGB photoionization to the one observed
\citep[see][for equivalent quantities]{bin94,cid11}. 
The latter is optionally corrected for intrinsic extinction, assuming this to
be equal to the extinction A$_V$ in the stellar component (cf K12 and G13). 
Since spectral fits imply a low ($\leq$0.3 mag) A$_V$ in most cases, this 
correction typically has a weak effect on $\tau$. We preferred to not base corrections of the
$\tau$ ratio on nebular extinction estimates since these are consistent with
A$_V$ within their uncertainties. 

We note that state-of-the-art SSP models imply that the \lyc\ photon rate per 
unit mass from pAGB stellar populations of nearly-solar metallicity
(0.008$\la Z \la$0.03) is almost independent of age, metallicity, and star
formation history \citep[e.g.][G13]{cid11}.
However, substantial uncertainties stem from the fact that existing models differ from 
one another by a factor $\sim$2 in the mean \lyc\ output they
predict for the pAGB stellar component \citep[][see also, e.g., Brown et al. 2008 and
  Woods \& Gilfanov 2013 for a discussion related to this subject]{cid11}.
These theoretical uncertainties presumably prevent a determination of the $\tau$
ratio to a precision better than within a factor of $\sim$2 from currently
available SSP models.

Our analysis in Sects.~\ref{r_vs_BPT} and \ref{r_vs_i} uses two complementary data sets:
i) single-spaxel (\sisp) determinations from fits with an absolute deviation
$\mid\!\!O_{\lambda}-M_{\lambda}\!\!\mid$/$O_{\lambda}$$\leq 2.6$ (cf K12), where
  $O_{\lambda}$ is the observed spectrum and $M_{\lambda}$ the fit.
These are typically restricted to the central, brightest part ($\mu\la$23 $g$ \sbb) of our sample ETGs.
ii) The average of all single-spaxel determinations within isophotal annuli (\iaa) 
adapted to the morphology of the (line-free) continuum between 6390 \AA\ and 6490 \AA\ (cf K12). 
These data, which are to be considered in a \emph{statistical sense}, go $\ga$2~mag fainter, 
allowing study of the azimuthally averaged properties of the \wism\ in the ETG periphery.
%
% ::::::::::::::::::::::::::::::::: Fig. 1: radial distribution of BPT ratios, EWs and the tau ratio :::::::::::::::
\begin{figure}
\begin{picture}(8.6,13.8)
\put(0.3,0.4){\includegraphics[width=0.326\textwidth, viewport=20 30 520 200]{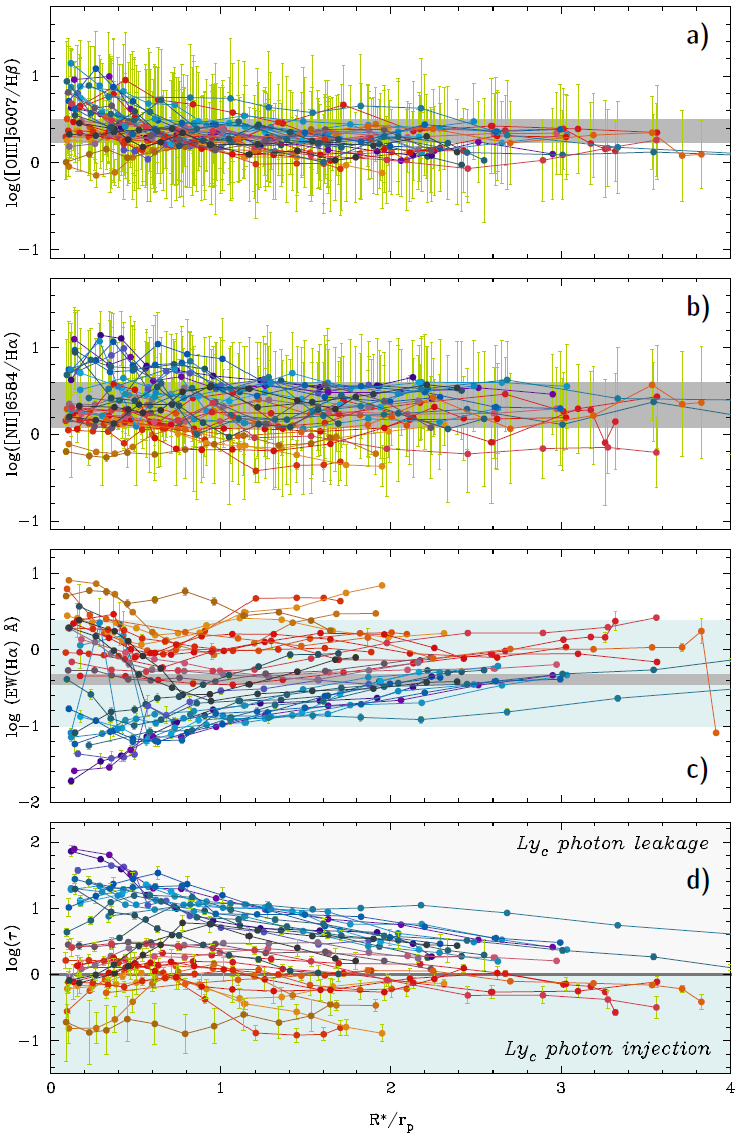}}
%\put(-0.4,9.9){\includegraphics[width=0.37\textwidth, viewport=20 30 520 200]{fig1a.pdf}}
%\put(-0.4,6.6){\includegraphics[width=0.37\textwidth, viewport=20 30 520 200]{fig1b.pdf}}
%\put(-0.4,3.3){\includegraphics[width=0.37\textwidth, viewport=20 30 520 200]{fig1c.pdf}}
%\put(-0.4,0.0){\includegraphics[width=0.37\textwidth, viewport=20 30 520 200]{fig1d.pdf}}
%
%\PutLabel{8.3}{13.2}{\vcap a)}
%\PutLabel{8.3}{9.9}{\vcap b)}
%\PutLabel{8.3}{4.3}{\vcap c)}
%\PutLabel{8.3}{2.95}{\vcap d)}
\end{picture}
\caption[]{{\sl From top to bottom:} \lo3hb, \ln2ha, $\log$(EW(\ha)), and $\log$(\tauha_ext) vs
normalized photometric radius \rr/\rPetr.
The gray shaded areas in panels a\&b mark the mean and $\pm$1$\sigma$ of the respective quantity,
and in panel~c the mean EW(\ha) for \rr$\geq$\rPetr\ (0.43$\pm$0.65 \AA).
The light-blue area in panel~c depicts the range in EW(\ha) that can be accounted for 
by pAGB photoionization models (0.1--2.4~\AA). 
The color assigned to each ETG is related to its <$\tau$> (cf text and Fig. \ref{fig:tau2})
in ascending order, from orange to violet, and is identical in all figures.}
\label{fig:r_vs_BPT}
\end{figure} 
% ::::::::::::::::::::::::::::::::::::::::::::::::::::::::::::::::::::::::::::::::::::::::::::::::::::::::::::::::::

% ::::::::::::::::::::::::::::::::: Fig. 2: tau vs I(Ha), EW(Ha), [OIII]/Hb and [NII]/Ha :::::::::::::::::::::::::::
\begin{figure*}
\begin{picture}(16.4,6.4)
\put(0.4,0.4){\includegraphics[width=0.326\textwidth, viewport=20 30 520 200]{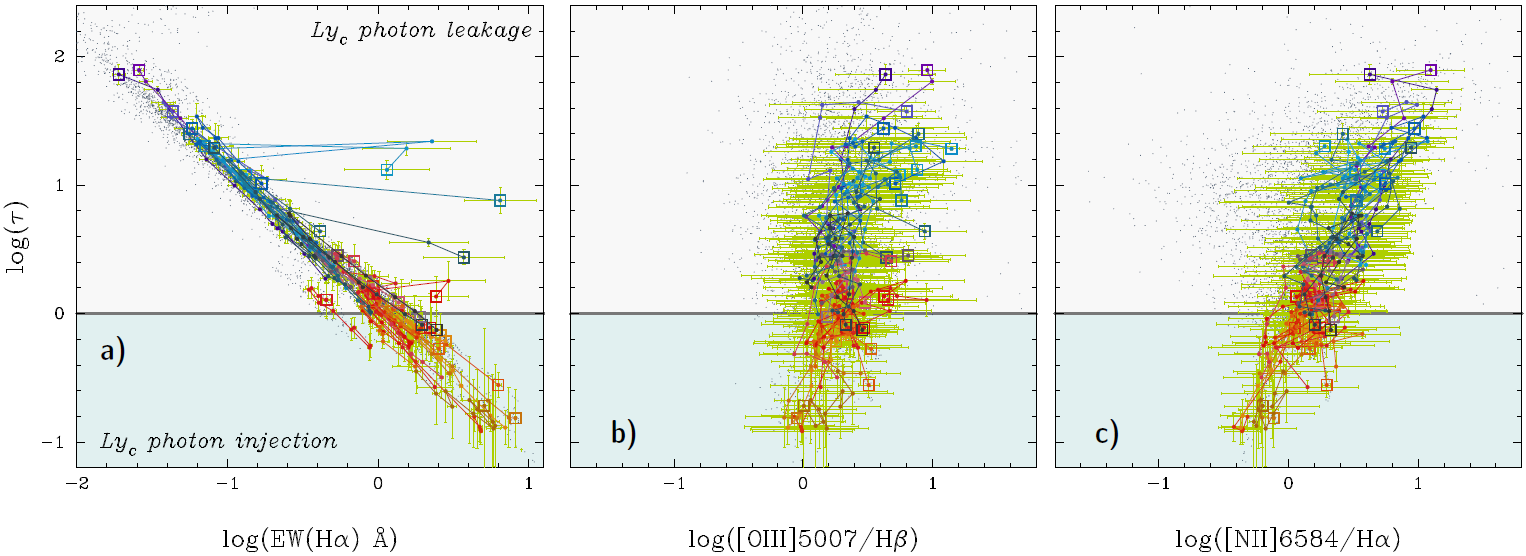}}
%\put(-2.0,0){\includegraphics[width=0.34\textwidth, viewport=20 30 520 200]{fig2a.pdf}}
%\put(4.0,0){\includegraphics[width=0.34\textwidth, viewport=20 30 520 200]{fig2b.pdf}}
%\put(9.9,0){\includegraphics[width=0.34\textwidth, viewport=20 30 520 200]{fig2c.pdf}}
%\PutLabel{1.1}{2.3}{\lcap a)}
%\PutLabel{7.3}{1.3}{\lcap b)}
%\PutLabel{13.2}{1.3}{\lcap c)}
\end{picture}
\caption[]{{\rem a}--{\rem c)} $\log$(EW(\ha)), \lo3hb\ and \ln2ha\ vs $\log$(\tauha_ext) for our sample ETGs. 
Open squares mark the central value for each \iaa\ profile (interconnected symbols), and dots show 
\sisp\ determinations.}
\label{fig:tau1}
\end{figure*} 
% ::::::::::::::::::::::::::::::::::::::::::::::::::::::::::::::::::::::::::::::::::::::::::::::::::::::::::::::::::

% ::::::::::::::::::::::::::::::::: Fig. 3 :::::::::::::::::::::::::::::::::::::::::::::::::::::::::::::::::::::::::
\begin{figure}
\begin{picture}(8.6,6.0)
\put(0.1,0.3){\includegraphics[width=0.334\textwidth, viewport=20 30 520 200]{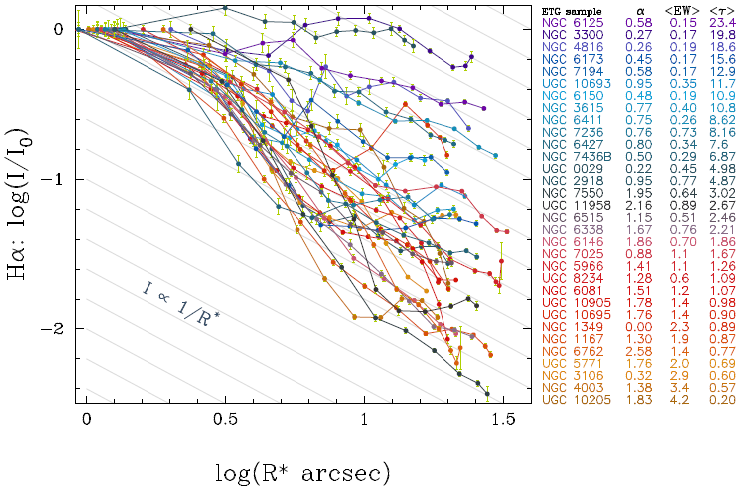}}
%\put(0,0){\includegraphics[width=0.33\textwidth, viewport=20 30 520 200]{fig3.pdf}}
\end{picture}
\caption[]{Normalized \ha\ intensity vs $\log(R^{\star})$ for our sample 
ETGs, based on \iaa\ determinations.
The diagonal lines correspond to a power-law intensity drop-off of the form 
$\log(I/I_0) \propto -\alpha\cdot log(R^{\star})$, with $\alpha=1$. 
The right-hand side table lists the power-law slope $\alpha$ and the radially averaged EW(\ha) and $\tau$ for each ETG.
}
\label{fig:tau2}
\end{figure} 
% ::::::::::::::::::::::::::::::::::::::::::::::::::::::::::::::::::::::::::::::::::::::::::::::::::::::::::::::::::
% ==========================================================================
\subsection{Radial behavior of emission-line diagnostics \label{r_vs_BPT}}
% ==========================================================================
Figures~\ref{fig:r_vs_BPT}a\&b show the diagnostic \lo3hb\ and \ln2ha\ line ratios 
for our sample ETGs as a function of the photometric radius \rr, normalized to the SDSS $r$ band Petrosian\_50 radius \rPetr.
The profiles are based on \iaa\ determinations, with green error bars illustrating the 
1$\sigma$ dispersion (typically $\sim$0.4 dex) of single-spaxel data points within each annulus. 
All galaxies show LINER-specific \citet[][BPT]{bpt81} ratios out to their periphery,
with weak (if any) gradients solely within their central part (\rr$\la$\rPetr).
The mean ratios for our sample (shaded regions) were determined to be 0.37$\pm$0.13 for \lo3hb\
and 0.34$\pm$0.26 for \ln2ha, with a standard deviation about the 
mean $\sigma_{\rm N}$ of 0.02 and 0.05.

The EW(\ha) profiles (panel~c) reveal a more complex pattern. For \rr$\ga$\rPetr, most data points 
fall between 0.1~\AA\ (\pAGBmin) and 
2.4~\AA\ (\pAGBmax), 
in the range of predictions from \pAGBp\ models
\citep[e.g.,][G13]{bin94,cid11}, whereas at smaller radii the sample seems to diverge
into a lower ($\la$\pAGBmin) and upper ($\ga$\pAGBmax) branch.

The $\tau$ ratio profiles (panel~d) include correction for intrinsic extinction, with
vertical bars illustrating the effect that neglecting it would have.
The reference line at $\log$(\tauha_ext)=0 corresponds to an equilibrium state where the 
\lyc\ photon output from pAGB stars balances the observed \ha\ luminosity. 
Values below ($\log$(\tauha_ext)$<$0) or above ($\log$(\tauha_ext)$>$0) that line
imply, in the first case, \lyc\ photon injection by an additional source (e.g., star formation, 
AGN, shocks) and, in the second, \lyc\ photon escape with a 
{\sl photon leakage fraction} \plf\ = 1-\tauha_ext$^{-1}$.

Setting a tentative division line at a radially averaged <$\tau$>=2, we can see that 
our ETG sample segregates into two groups.
In the first one (type~i; <$\tau$>\,$<$\,2, 14 ETGs), 
the $\tau$ ratio shows little dependence on radius, with individual data points 
deviating in most cases by no more than 0.3~dex from the equality line.
This suggests a moderate \lyc\ leakage (\plf$\leq$0.5) and/or dominant contribution 
of \pAGBp\ to the excitation of the \wism.
In the second group (type~ii; <$\tau$>\,$\geq$\,2, 18 ETGs), the \plf\ is typically very large 
($\ga$0.9) within \rPetr, and far from negligible ($\ga$0.6) even in the galaxy periphery.
As is apparent from panel~c, these two groups differ in their EW(\ha), 
with radially averaged values <EW> of
1.82$\pm$1.04 \AA\ ($\sigma_{\rm N}$=0.28 \AA) and  
0.41$\pm$0.25 \AA\ ($\sigma_{\rm N}$=0.06 \AA).
Another salient feature is that EW profiles of type~i ETGs are nearly constant
beyond $\sim$\rPetr/2, whereas those of type~ii ETGs show a tendency toward 
a smooth, monotonic increase out to their periphery. 

Figures~\ref{fig:tau1}a-c display projections of some quantities of interest onto \tauha_ext.
Unsurprisingly, both \sisp\ and \iaa\ data delineate a trend toward decreasing EW(\ha)
with increasing \tauha_ext, with type~i and type~ii ETGs populating, respectively, 
the lower and upper parts of a continuous sequence (panel a). 
This trend is also reflected on a relation
log<$\tau$>=(0.23$\pm$0.04)--(1.36$\pm$0.09)$\cdot$log<EW(H$\alpha$)> for our sample
(cf right-hand side list in Fig.~\ref{fig:tau2} for the <$\tau$> and <EW>
of individual ETGs).
On the \lo3hb\ vs log($\tau$) plane (panel b), the two ETG groups differ only marginally from one 
another (\lo3hb\ of 0.29$\pm$0.11 and 0.43$\pm$0.12), while a weak trend 
toward increasing \ln2ha\ with log($\tau$) is apparent from panel~c 
(0.11$\pm$0.13 and 0.52$\pm$0.17 for type~i and type~ii ETGs, respectively).
%
% =========================================================================================
\subsection{Radial intensity distribution of nebular emission \label{r_vs_i}}
% =========================================================================================
The radial \ha\ intensity profiles in Fig.~\ref{fig:tau2} indicate  
that faint \ige\ is present over nearly the entire optical 
extent of our sample ETGs.
From the Abel integral equation \cite[see, e.g.,][for a discussion and
solutions for various intensity profiles]{P96a} 
it follows that, for an isotropically emitting 
spheric-symmetric volume, an intrinsic luminosity density distribution $l(r)$ scaling as
$\propto r^{-2}$ would be projected onto a power-law intensity profile of the form 
$\log(I/I_0) \propto -\alpha\log(R^{\star})$ with $\alpha=1$. 
On the simplifying assumption that the \lyc\ output from a putative AGN is  
internally reprocessed into \ige\ with a $l(r) \propto r^{-2}$, 
one can invoke the $\alpha$ inferred from \ha\ profile fitting
as a minimum consistency check for the AGN illumination hypothesis.
The mean $\alpha$ for our sample (1.09), obtained for 
\rr$\geq$3\farcs7 (the effective FWHM resolution of CALIFA IFS cubes) is indeed 
consistent with it and close to the value deduced by
\citet[][$\alpha$=1.28]{YanBlanton2012} from comparison of two-aperture spectroscopic data.
Nevertheless, the large standard deviation in the derived slopes ($\sigma=0.67$) 
argues against a \emph{universal} power-law index $\alpha\approx1$ 
for the \ige\ intensity drop-off in ETGs.
It is interesting though that comparison of Figs.~\ref{fig:r_vs_BPT}\&\ref{fig:tau2}
suggests a tendency for type~ii ETGs having shallower \ha\ profiles ($\alpha=0.85\pm0.56$; $\sigma_{\rm N}=0.13$) 
than type~i ETGs ($\alpha=1.40\pm0.67$; $\sigma_{\rm N}=0.18$).
%
% ==========================================================================
\section{Discussion and conclusions \label{disc}}
% ==========================================================================
Summarizing the evidence from Sect.~\ref{meth}, the ETGs studied here form a broad, continuous 
sequence with respect to their \ha, EW(\ha), and $\tau$ profiles.
Adopting a radially averaged $\tau$ ratio cutoff of <$\tau$>=2, we tentatively subdivide 
our sample into two groups:
Typical properties of type~i ETGs are a rather steep \ha\ drop-off ($\alpha>1$), 
nearly constant EWs of $\ga$1~\AA\ beyond \rPetr, and a <$\tau$> close to unity (0.3\dots2). 
Type~ii ETGs display shallower \ha\ profiles ($\alpha\!<\!1$),
overall very low ($\la$\pAGBmin\dots 0.5~\AA), outwardly increasing EWs, and a
large (up to $\sim$20) <$\tau$>.
Despite a difference of almost 2~dex in their nuclear $\tau$ ratios, these two groups
differ little (by $\la$0.4~dex) in their mean \lo3hb\ and \ln2ha\ BPT ratios, 
which in either case are characteristic of LINERs and, within their uncertainties 
($\sim$0.4\ dex), are radially constant.
In our ETG sample, 64\% of the S0 galaxies fall into the type~i group, and 
78\% of the E galaxies fall into the type~ii group.
Clearly, the classification proposed here is only indicative 
and needs to be refined, both by obtaining better statistics and through 
a quantitative comparison with other ETG properties: 
of these, the X-ray luminosity and temperature, the $\alpha_4$ 
and $(v/\sigma)_{\star}$ parameter, and the star formation history are all 
being actively investigated.

As far as type~i ETGs are concerned, various lines of evidence from this study suggest, 
in line with a substantial body of previous work \citep[e.g.][K12, among others]{sar10,ani10,YanBlanton2012}, 
that \pAGBp\ is the main driver of extended \ige, with nonthermal sources only
being potentially important in nuclei:
\rem{a)} First, \ige\ is not confined only to the nuclear regions but is
extended out to \rr$\sim$\,2--4\rPetr,  i.e. is co-spatial with the pAGB stellar background.
\rem{b)} Second, radial \ha\ profiles rule out a power-law intensity drop-off 
with a \emph{universal} slope $\alpha\approx1$, as a possible signature of a 
\emph{dominant} AGN contribution to the excitation of the \wism.
\rem{c)} Third, the EW(\ha) is nearly constant beyond $\sim$\rPetr,
pointing to a causal relationship between \ige\ and the projected stellar 
surface density \sstar\ along the line of sight.
This is a plausible expectation from the \pAGBp\ scenario, further supported 
by the quantitative agreement between predicted and observed EWs, 
as well as the narrow range in $\tau$ ratios ($\simeq1$).
These presumably suggest that type~i ETGs contain a sufficient 
amount of \wism\ being well mixed with stars, to justify case~B 
recombination as a first-order approximation.
Conversely, it is not immediately apparent how AGN or shock excitation  
alone could lead to the remarkable fine tuning between EW(\ha) and \sstar\ 
over $\ga$2\ dex in \sstar. 

The emerging picture for type~ii ETGs appears more complex.
The \tauha_ext\ ratios inferred for these systems 
imply at face value that the bulk (70\%\dots$\ga$90\%) of the \lyc\ output 
from hot pAGB stars (\emph{consequently, from any other discrete or diffuse ionizing source}) 
escapes into the intergalactic space without being reprocessed into \ige.
Admittedly, \plf\ estimates depend on the pAGB mass inferred
from spectral synthesis models. These are known to suffer from degeneracies, the  
amplitude, topology, and systematics of which remain almost uncharted territory.
One might argue that the large number of fits per galaxy (up to $\sim$6800, using
two SSP libraries) permits eliminating uncertainties, at least as far as \iaa\ 
determinations are concerned. However, this argument would only apply if errors in 
spectral synthesis were uncorrelated and nonsystematic.\\
Nevertheless, in the specific context of type~ii ETGs, an essentially 
model-independent argument for extensive \lyc\ leakage comes from the 
virtual absence of \ige, despite a sizeable ionizing photon budget.
Quantitatively, for the \lyc\ escape interpretation to become untenable,
\ha\ fluxes in type~ii ETGs would need be revised upward 
by up to two orders of magnitude, which can be excluded by any reasonable error budget.

The $\tau$ profiles of these ETGs consistently point toward 
a low, radially dependent volume-filling factor \ff\ and/or density for the \wism.
A further hint in the same direction comes from their \emph{positive} EW gradients:
For a spheric-symmetric volume, these in connection 
with shallow ($\alpha\!\!<\!\!1$) \ha\ profiles are only reproducible when the \wism\ 
luminosity density is monotonically decreasing toward the center.
Alternatively, a feature predicted (though not observed) by \citet[][see their Fig.~10]{sar10} 
are positive EW gradients in a spherical stellar host reprocessing its 
pAGB \lyc\ output within a \emph{planar} gas configuration.
Expanding the considerations by these authors, the high \plf's and outwardly 
increasing EWs of type~ii ETGs might be reconcilable for a geometry that involves an 
oblate distribution of tenuous/clumpy gas within a spherical stellar host.
Evidently, such a geometry would \emph{per se} imply pAGB UV photon escape, further reinforcing 
our interpretation.

Regardless of the 3D distribution of the \wism, its high porosity 
and/or low \ff\ call into question the importance of shock 
excitation in type~ii ETGs. If the \wism\ is primarily composed of 
compact cloudlets of radius $r_{\rm c}$, then their large mean-free-path 
\citep[(2/3)\,$r_{\rm c}$/\ff, e.g.,][]{JogSolomon1992} 
would act toward reducing the efficiency of energy dissipation 
via cloud-cloud collisions and shocks.

Given our findings, it is important to ask whether the ``weak'' AGN
interpretation of the optical emission lines is compelling anymore.
In the presence of extensive \lyc\ leakage, emission-line intensities
and EWs in type~ii ETG nuclei are lowered by factor between $\sim$10
and $\la$100.
Consequently, the presence of weak \ige\ in these systems is not in
itself proof of a ``weak'' (sub-Eddington accreting) AGN. 
In fact, the importance of \lyc\ photon escape, which heretofore has not been
investigated in detail, may be a key insight into resolving one of the
longstanding enigmas in AGN/LINER research. It offers an \emph{ansatz} for
reconciling the fact that many ETGs with prominent signatures of strong
AGN activity in radio continuum and/or X-ray wavelengths merely show weak
(LINER) optical AGN signatures.

In addition, the relative distribution of gas compared to the stars is an important issue. 
While in a thin, face-on disk, a nuclear EW of, say, $\la$10 \AA\ can safely be 
regarded as evidence of faint \ige\ (and a weak AGN), this is not necessarily 
the case for a \emph{triaxial} ETG, where the \ige-emitting gas volume may have 
a more limited extent than the stellar component.
In the latter case, nuclear EWs are effectively lowered by the high-surface brightness screen
of background and foreground stars along the optical path. 
In conjunction with the \lyc\ photon escape, this line-of-sight dilution of the nuclear
EWs will conspire to create an observational selection effect, favoring optical
detection of AGN activity in oblate, face-on ETGs with atypically low \plf's.

Arguably, one of the most surprising results of this study are the similar 
mean BPT ratios of type~i and type~ii ETGs, despite having substantial differences 
in their \wism\ characteristics.
Perhaps the luminosity-weighted emission-line ratios projected along the line of
sight ``saturate'' into the LINER regime for a broad range of \wism\
distributions and characteristics (i.e. combinations of differing
\ff, covering fractions, densities, ionizing photon mean free-path,
and ionization parameters), becoming degenerate for ETGs.
Circumstantial support for this hypothesis comes from radiation
transfer models by \cite{Ercolano2012}, who report that a subset of the
\emph{projected} emission line ratio determinations in a 3D model of the
{\sl Pillars of Creation} can mimic LINER characteristics in classical
(1D) BPT diagrams.  Clearly, detailed 3D radiative transfer modeling
of the \wism\ in ETGs, including nonequilibrium ionization
effects \citep[e.g.,][]{deAvillezBreitschwerdt2012}, are important for
understanding the nature of \wism\ in ETGs.  High-quality IFS data,
such as those from CALIFA, will no doubt provide crucial observational
constraints on these next-generation models.

\begin{acknowledgements}
We thank the anonymous referee for valuable suggestions.
This paper is based on data from the Calar Alto Legacy Integral Field
Area Survey, CALIFA (http://califa.caha.es), funded by
the Spanish Ministery of Science under grant ICTS-2009-10, and the
Centro Astron\'omico Hispano-Alem\'an. 
PP is supported by Ciencia 2008 Contract, funded by FCT/MCTES 
(Portugal) and POPH/FSE (EC), and JMG by a Post-Doctoral 
grant, funded by FCT/MCTES (Portugal) and POPH/FSE (EC).
% ::::::::::::::::::::::::::::::::::::::::::::::::::::::::::::::::::::::::::::::::::::::::::::
% 
PP\&JMG acknowledge support by the Funda\c{c}\~{a}o para a Ci\^{e}ncia e a Tecnologia (FCT)
under project FCOMP-01-0124-FEDER-029170 
(Reference FCT PTDC/FIS-AST/3214/2012), funded by FCT-MEC (PIDDAC) and FEDER (COMPETE).
IM acknowledges support from Spanish grant AYA2010-15169 and the Junta de
Andalucia through TIC-114 and the Excellence Project P08-TIC-03531.
J.F-B. from the Ram\'on y Cajal Program, grants AYA2010-21322-C03-02 and
AIB-2010-DE-00227 from the Spanish Ministry of Economy and Competitiveness (MINECO), as
well as from the FP7 Marie Curie Actions of the European Commission, via the Initial
Training Network DAGAL under REA grant agreement n$\circ$ 289313.
% ::::::::::::::::::::::::::::::::::::::::::::::::::::::::::::::::::::::::::::::::::::::::::::
The {\sc starlight} project is supported by the Brazilian agencies CNPq, CAPES, and FAPESP. 
PP has enjoyed inspiring discussions on the effects of \lyc\ photon leakage
with Prof. Nils Bergvall (Uppsala University). 
We benefited from stimulating discussions with several members of the CALIFA collaboration.
This research made use of the NASA/IPAC Extragalactic Database (NED) which is
operated by the Jet Propulsion Laboratory, California Institute of
Technology, under contract with the National Aeronautics and Space
Administration.
\end{acknowledgements}
\end{document}